\documentclass[doublecol]{epl2}

\usepackage{amsmath}
\usepackage{amssymb}

\title{Energy distributions and effective temperatures in the packing of elastic sheets}
\shorttitle{Energy distributions and effective temperatures in the packing of elastic sheets} 

\author{S. Deboeuf \and M. Adda-Bedia \and A. Boudaoud}
\shortauthor{S. Deboeuf \etal}

\institute{
Laboratoire de Physique Statistique de l'Ecole Normale Sup\'erieure, CNRS UMR 8550 - 24 rue Lhomond, 75005 Paris, France
}


\abstract{
The packing of elastic sheets is investigated in a quasi two-dimensional experimental setup: a sheet is pulled through a rigid hole acting as a container, so that its configuration is mostly prescribed by the cross-section of the sheet in the plane of the hole. The characterisation of  the packed configuration is made possible by using refined image analysis. The geometrical properties and energies of the branches forming the cross-section are broadly distributed. We find distributions of energy with exponential tails. This setup naturally divides the system into two sub-systems: in contact with the container and within the bulk.
While the geometrical properties of the sub-systems differ, their energy distributions are identical, indicating 'thermal' homogeneity and allowing the definition of effective temperatures from the characteristic scales of the energy distributions.
}

\begin{document}

\maketitle

\section{Introduction}

The challenges raised by out-of-equilibrium systems are exemplified by granular materials~\cite{jaeger96} and glasses~\cite{bouchaud98,debenedetti01}, featuring complex energy landscapes and aging.  Energy flow, thermal equilibration, and the statistical properties of energy in such systems can be characterised by various effective temperatures~\cite{edwards89,cugliandolo97a,bertin06}; however, previous experimental studies~\cite{grigera99,cugliandolo99,bellon01,song05,wang06,jabbari07} only measured a temperature based on the ratio between fluctuations and response of the system.
Here we present experiments on a macroscopic out-of-equilibrium system, namely the packing of elastic sheets into quasi two-dimensional containers~\cite{boue06} and focus on the statistical properties of the configurations. We measure the distributions of geometrical and energetic properties and show thermal homogeneity within the system although its geometrical properties are not uniform, enabling the definition of effective temperatures from the distributions of energy. Thus we obtain a macroscopic experimental system that could be used to test out-of-equilibrium statistical physics. Our results bear on the packing of flexible structures such as elastic rods~\cite{donato03,katzav06,boue07,stoop08}, crumpled paper~\cite{matan02,blair05,sultan06,balankin06}, folded leaves in buds~\cite{kobayashi98}, chromatin in cell nuclei~\cite{kleckner04} or DNA in viral capsids~\cite{purohit05,katzav06}.

At equilibrium, systems with a large number of degrees of freedom are characterised by a single temperature $T$. On the one hand, the energy of one degree of freedom follows Boltzmann's distribution, the mean energy being proportional to $T$. On the other hand, $T$ might be measured using the fluctuation-dissipation theorem (FDT), relating fluctuations of an observable to its response to an external field. By analogy, two main effective temperatures were introduced for systems out of equilibrium. The approach of Edwards~\cite{edwards89} amounts to the replacement of $T$ by an effective temperature in the distribution of energies; it can be extended to intensive thermodynamic parameters associated with global conserved quantities~\cite{schroter05,bertin06,lechenault06,aste07}. The generalisation of the FDT~\cite{cugliandolo97a} gives another effective temperature, which can be measured~\cite{grigera99,cugliandolo99,bellon01,song05,wang06,jabbari07,berthier02}. In many models, Edwards' and FDT temperatures are equal~\cite{barrat00,makse02,ono02,dean03} or proportional~\cite{shokef06}.  An experimental measurement of Edwards' temperature seems to be lacking as it is difficult to obtain energy distributions. Here we measure energy distributions in the packing of elastic sheets.

\begin{figure}
\begin{minipage}{\linewidth}
\textbf{a)} \\ \centerline{\includegraphics[width=.55\linewidth]{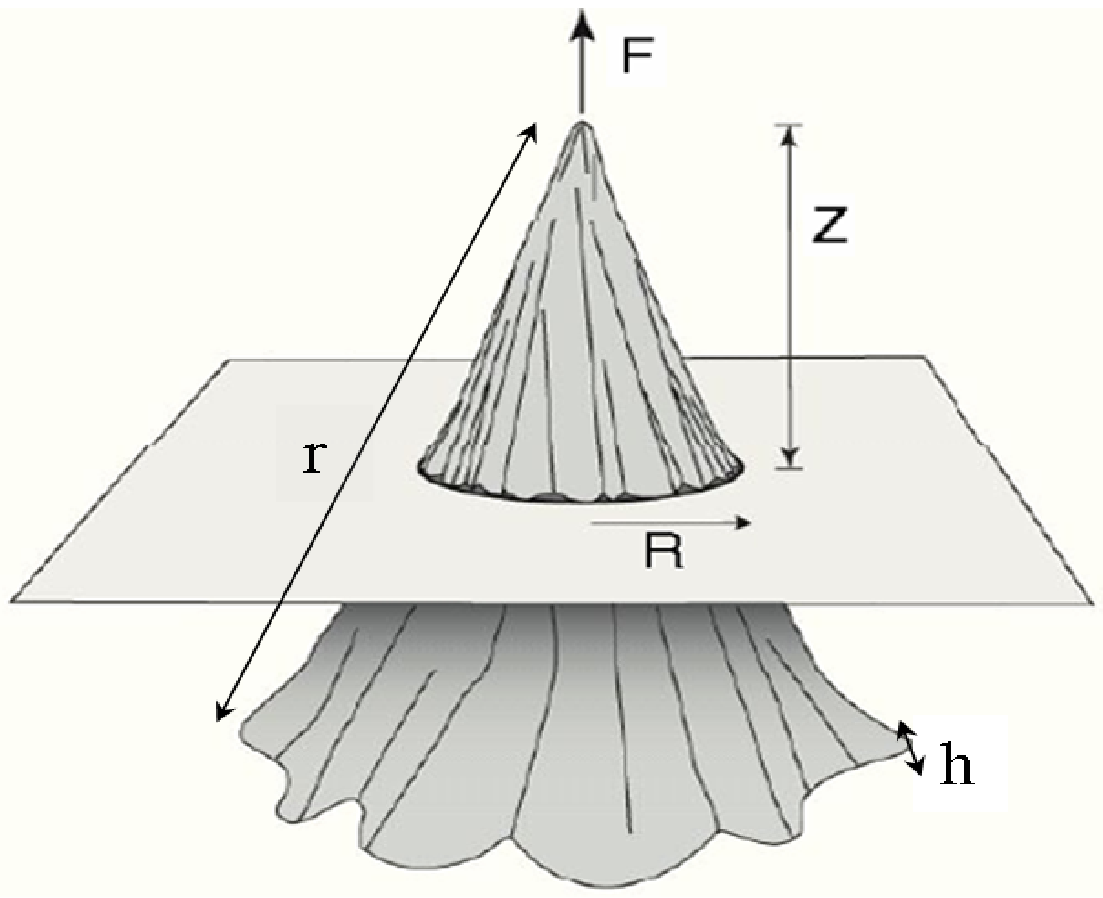}}
\end{minipage} \\
\begin{minipage}{\linewidth}
\textbf{b)} \\ \centerline{\includegraphics[width=.4\linewidth]{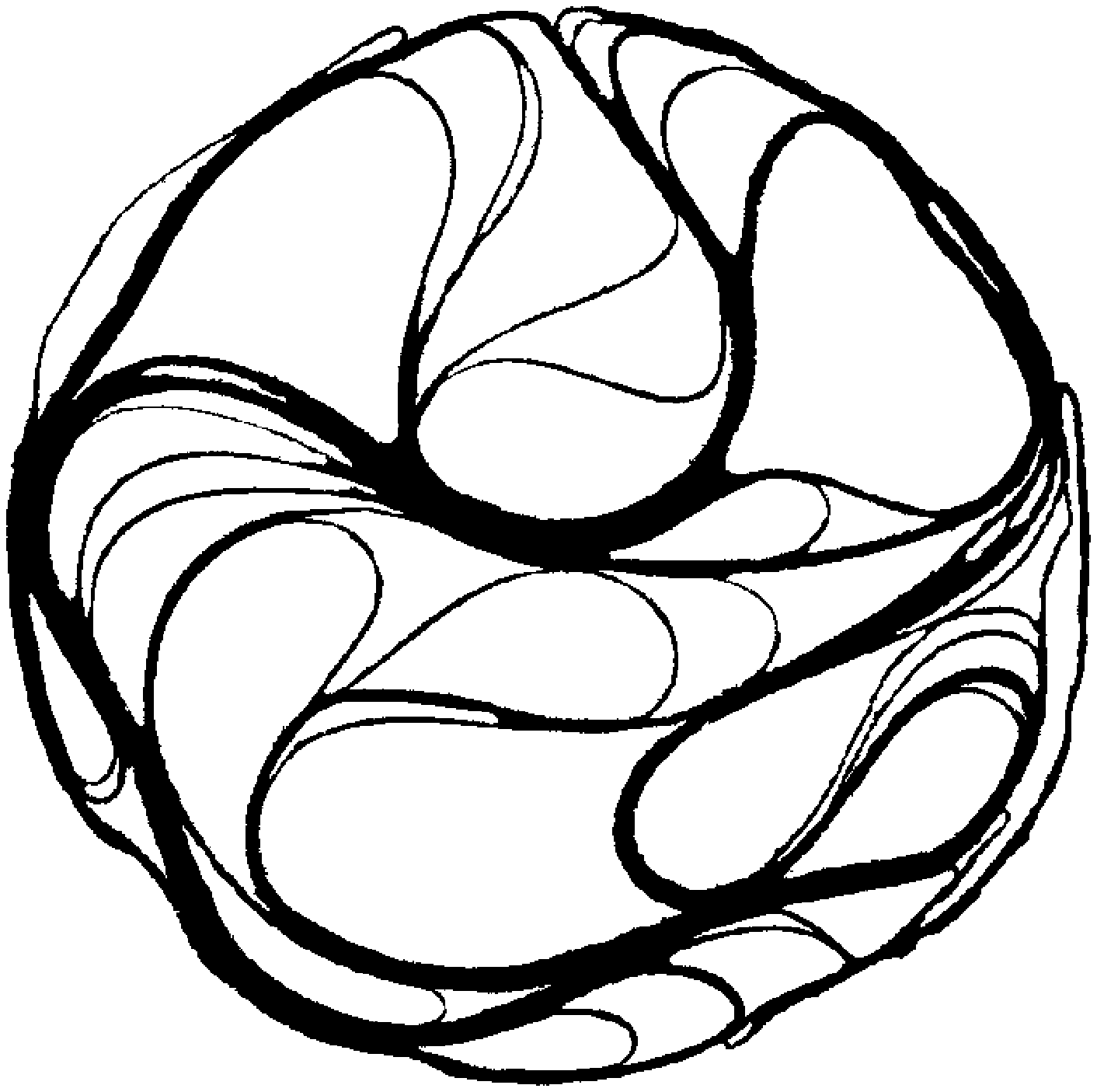}}
\end{minipage} \\
\begin{minipage}{\linewidth}
\textbf{c)} \\ \centerline{\includegraphics[width=.4\linewidth]{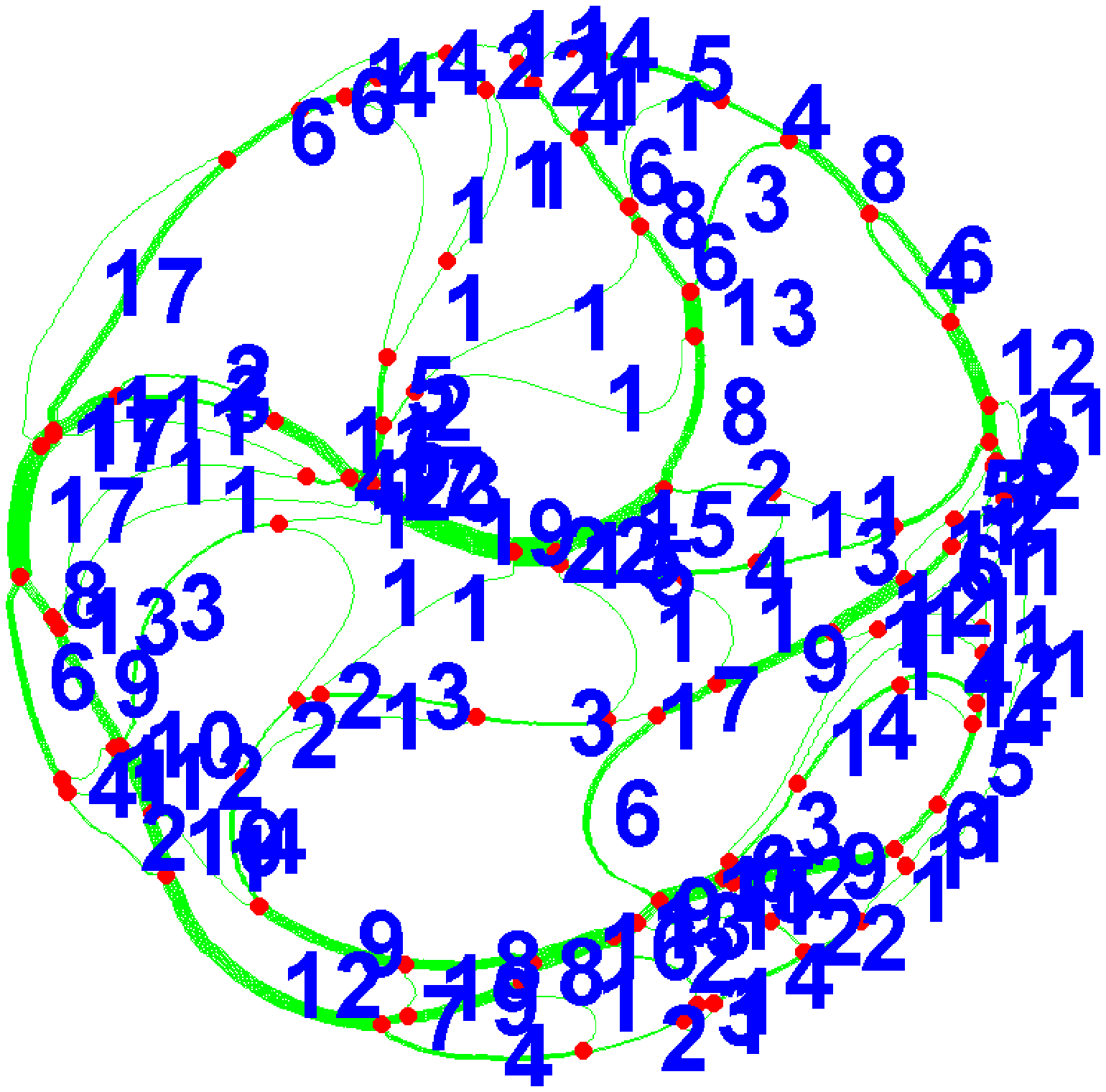}}
\end{minipage} \\
\begin{minipage}{\linewidth}
\textbf{d)} \\ \centerline{\includegraphics[width=.6\linewidth]{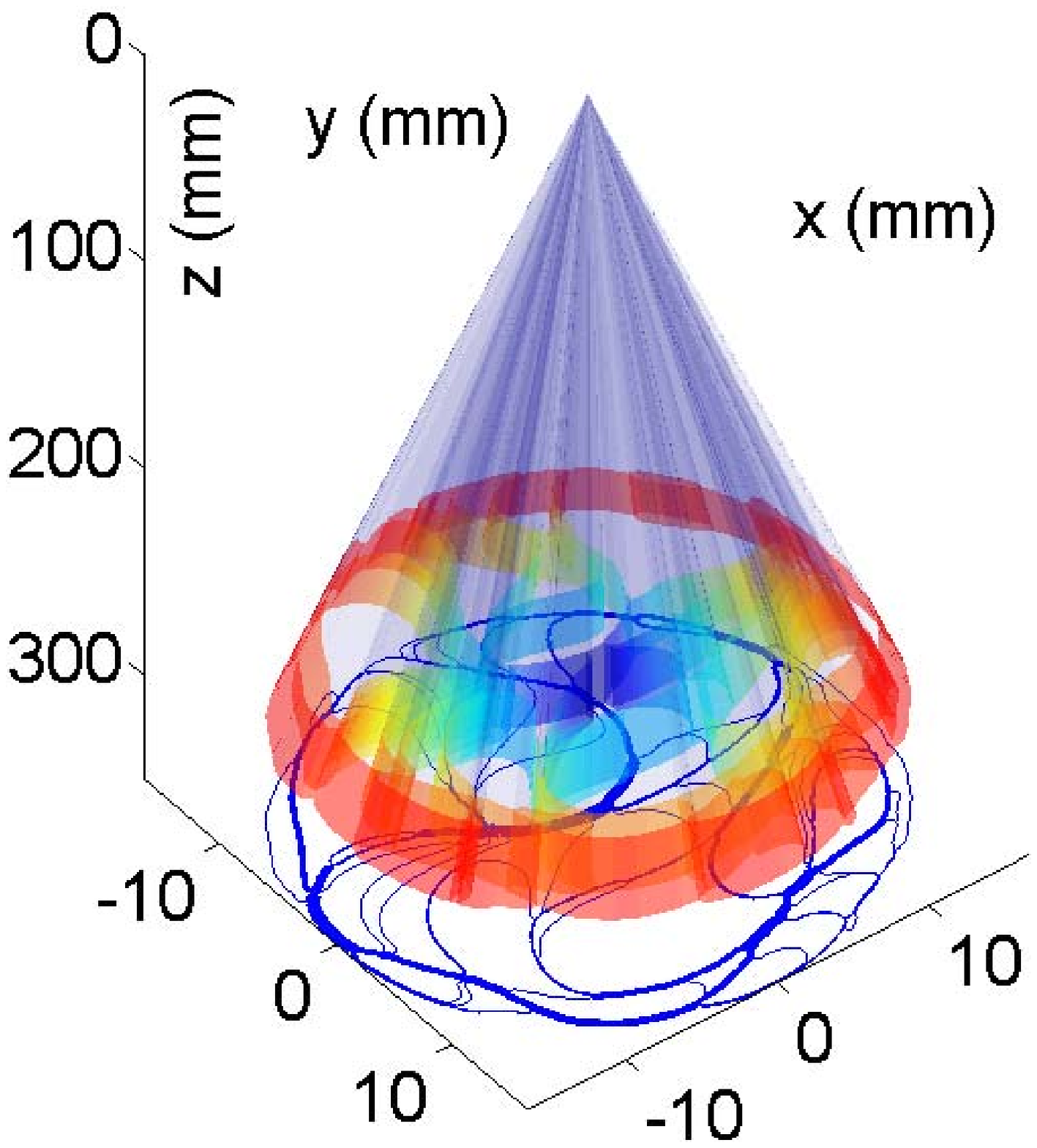}}
\end{minipage} 
\caption{The experiment. \textbf{a)} Sketch of the set-up showing the radius of the sheet $r$, its thickness $h$, the radius of the hole $R$ and the control parameter $Z$; the force $F$ is measured with a dynamometer. \textbf{b)} Thresholded picture of a horizontal cross-section of a configuration from set $i$ of experiments. \textbf{c)} Analysed cross-section, showing the existence of multi-branches stacks delimited by two junction points. The number of branches is indicated near each stack. \textbf{d)} 3D reconstruction of the same configuration, assuming exact self-similarity of shape. }
\label{setup}
\end{figure}

\section{Experimental set-up}

Fig.~\ref{setup}\textbf{a} represents the experimental set-up, as introduced in~\cite{boue06}, which was inspired by the study of single d-cones~\cite{chaieb98}. We use circular polyester (polyethylene terephtalate) sheets of Young's modulus measured as $E=5$~GPa, density $1.4$~g/cm$^3$, various radii $r\sim30$~cm and thicknesses $h\sim100~\mu$m (see Table~\ref{tab1}). At each realisation, a sheet is pulled from its centre through a circular rigid hole of radius $R\sim20$~mm. The values of the parameters ($r,h,R$) for each set of experiments are given in Table~\ref{tab1}. The hole is machined through a Plexiglas plate and its edges are rounded to form a toroidal convex shape, to avoid damaging the sheet. The center of the sheet is pierced and fixed to a dynamometer by means of a threaded mount of radius $0.8$~cm. The sheet is pulled at a velocity of $0.5$~mm/s, so that the distance $Z$, between the pulling point and the plane of the hole is our main control parameter. The measurement of the pulling force $F$ during the compaction directly yields the work injected in the system $W=\int_0^{Z} F\mathrm{d}z$. This injected energy serves to pack the sheet and is dissipated through friction. The coefficients of friction for polyester/Plexiglass and polyester/polyester were measured as $0.37$ and $0.30$, respectively.

The sheet might undergo two modes of deformation: bending and
stretching. As bending is favoured energetically, a self-similar
conical shape is expected~\cite{benamar97}, so that one
cross-section approximately prescribes the whole shape of the sheet.
A virtual cut across the sheet in the plane of the hole yields a
one-dimensional rod of length $2\pi Z$, that grows within a disk of
radius $R$ as $Z$ is increased. The experiment allows
\emph{isotropic} confinement to packing ratios $P$ as high as
$0.11$, where $P=2Zh/R^2$ is the ratio of cross-sectional area of
the sheet $2\pi Z h$ to the area of the hole $\pi R^2$.

\begin{table*}
 \begin{center}
  \begin{tabular}{l||l|l|l|l||l|l|l||l|l}
 & $h$\ ($\mu$m) & $r$\ (cm) & $B$\ (J) & $\kappa_c$\ (mm$^{-1}$)& $R$\ (mm) & $Z_{\mbox{\tiny m}}$\ (cm) &
$P$ & $\#\mathcal{R}$ & $\sum N_{br}$ \\ \hline
$i$&50 & 33 & $7\,10^{-5}$ & 0.54 & 16.5 & 30 & 0.11 & 33 & 16170 \\
$ii$&50 & 33 & $7\,10^{-5}$ & 0.54 & 22.5 & 30 & 0.06 & 16 & 5760 \\
$iii$&125 & 22 & $1\,10^{-3}$ & 0.24 & 27 & 19  & 0.07 & 16 & 1760 
  \end{tabular}
  \end{center}
  \caption{Material parameters for the sheets used in experiments: thickness $h$, radius $r$, bending stiffness $B$ and plastic threshold curvature $\kappa_c$; Control parameters: hole radius $R$, maximal pulling distance $Z_{\mbox{\tiny m}}$, and packing ratio $P=2hZ_{\mbox{\tiny m}}/R^2$; Total numbers of realisations $\#\mathcal{R}$ and of branches $\sum N_{br}$, on which statistical analyses are based.} 
  \label{tab1}
\end{table*}

In principle, configurations can be visualised from below. However
this turns out to be inconvenient as parts of the sheet assemble
into thick bundles and the edge of the sheet does not lie in a
single plane. Therefore we resort to a hot wire cutting tool to
obtain cross-sections for one value of the control parameter
$Z_{\mbox{\tiny m}}$ (given in Table~\ref{tab1}). With great care,
one obtains neat cuts without perturbing the configuration. The
cross-section is digitised with a scanner at a resolution of 50
pixels per mm. A thresholding results in a binary image, in which
empty spaces of surface area larger than $(10h)^2$ are kept, which
removes light noise from the raw image, as shown in Fig.~\ref{setup}\textbf{b}.
The binary image is skeletonized (reduced to a one pixel thick
skeleton); junction points are then defined as pixels with at least
three neighbours. Two neighbouring junction points delimit a stack of
branches in close contact. The next step is to determine the number
of branches in each of the $M$ stacks. The conservation of the
number of branches at each junction point yields $2M/3$ equations, because 3 stacks intersect at each junction point; the
remaining $M/3$ equations are found from the thickness of the stacks
in the binary image as follows. The heating by the cutting tool
thickens (about twice) a stack nonlinearly, which was calibrated by
separately cutting stacks of sheets. As a result, the smallest
sheet thickness used  ($h$=50~$\mu$m) corresponds to 5 pixels. We keep the $M/3$ stacks with
the best estimation of the thickness as given by the calibration.
The solution of the linear $M\times M$ system yields the number of
branches in each stack. We reopened a few configurations (5 per set
of experiments) and checked by counting the number of branches in
each stack: we found no error for sets $ii$ and $iii$, and an error of
$\pm 1$ branch in $20\%$ of the stacks for the more
compact set $i$. These errors are small thanks to the fact that the
number of branches in a stack is an integer. Thus, we obtain both the geometry
and the topology of the sheet
(Fig.\ref{setup}\textbf{c},\textbf{d}). 

When repeating the
experiment  with the same experimental parameters, a whole variety of
shapes is generated, which calls for a statistical approach and an ensemble analysis. 
We systematically performed and analysed three sets of experiments
with a number of realisations $\#\mathcal{R}\sim20$ 
(Table~\ref{tab1}). We detail in the following our experimental results.

\begin{figure}
\center{\includegraphics[width=.8\linewidth]{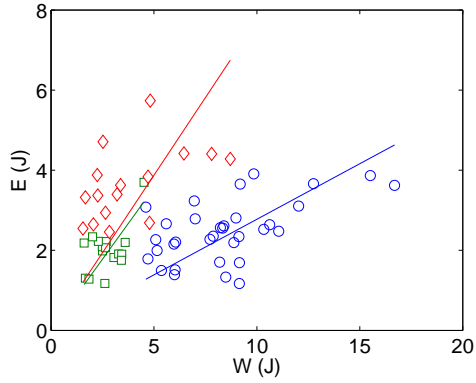}}
\caption{Total elastic energy $E$ (measured from the geometry according to Eq.~\ref{eqformenerg}) and
injected work $W$ (measured from the pulling force $F$) for the three sets of experiments: $i$ ($\circ$),
$ii$ ($\square$) and $iii$ ($\lozenge$). The straight lines are linear fits to
each set.} \label{temperato}
\end{figure}

\section{Total energy}

We first consider the global energetic quantities, injected work $W$
and elastic energy $E$, and their correlations. Assuming the shape of the folded sheet to be exactly self-similar, a cross-section prescribes the energy of the
whole sheet as follows~\cite{benamar97}. Using the polar coordinates $(\rho,\theta)$ on the
initially plane sheet, the branches, located in the plane
$\rho\simeq Z_{\mbox{\tiny m}}$, have a local curvature
$\kappa(\theta)$, and correspond to an angular sector on the sheet,
where the curvature is $c(\rho,\theta)=Z_{\mbox{\tiny
m}}\kappa(\theta)/\rho$, assuming the hole to be small ($R\ll
Z_{\mbox{\tiny m}}$). The bending energy $E$ of the whole sheet is
\begin{equation}
\label{eqformenerg} \frac{B}{2}\!
\int_{R_c}^{r}\!\int_{0}^{2\pi}\!\! c^2\, \rho\mathrm{d}\theta\,
\mathrm{d}\rho = \frac{BZ_{\mbox{\tiny m}}}{2}\
\ln\left(\frac{r}{R_c}\right)\! \int_{0}^{2\pi Z_{\mbox{\tiny m}}}\!
\kappa^2(t)\, \mathrm{d}t
\end{equation}
where we introduced $t=Z_{\mbox{\tiny m}}\theta$, the curvilinear
coordinate in the hole cross-section. The logarithmic prefactor
known for d-cones~\cite{benamar97} contains as cutoffs the radii of
the core of the cone $R_c$ and of the sheet $r$.  In actual experiments, the
self-similarity is not exact as some generators end below the mount.
This affects the logarithmic prefactor through
the effective value of $R_c$, which might lead to 
an error in the overall
multiplicative factor of order 1 in the estimation of the
total elastic energy. Here, we chose to estimate $R_c$ as the radius of the mount ($0.8$~cm). In order to account for plastic
softening of the sheets (as observed locally along a few scars), the quadratic $\kappa^2$ dependence of the
energy (Eq.~\ref{eqformenerg}) was replaced by a linear dependance
$\kappa_c(2\kappa-\kappa_c)$ for curvatures greater than the plastic
threshold $\kappa_c$ (Table~\ref{tab1}), measured as in~\cite{blair05}.  

Fig.~\ref{temperato} shows that the bending energy $E$ and the injected work $W$ are correlated. Indeed, for each set of experiments, $E$ is roughly proportional to $W$, showing that the stored elastic energy $E$ can be controlled with an external force. The unphysical values (mostly in set {\em iii}) such that $E>W$
can be mainly ascribed to the choice of $R_c$ as the radius of the mount; 
choosing, instead, $R_c$ of the order of the hole radius would shift all data below the line $E=W$.
Another possible source of bias comes from the estimation of  the energy of the very few branches with local high curvatures ($\kappa\gg\kappa_c$) that contribute significantly to the total energy. 
We stress however that these two sources of error do not affect the statistics of energy as discussed below.

Fig.~\ref{temperato} also shows
that the global quantities $E$ and $W$ fluctuate over the
realisations of a given set as the system explores its configurational space. 
Energy dissipation occurs by friction between layers and with the
container, and through discontinuous bifurcations~\cite{boue07} corresponding to reorganisations of the configurations when
the confinement is increased. The evolution of the overall slope of $E(W)$ suggests that 
the dissipated fraction of energy increases with confinement; indeed the more compact set {\em i} has the smallest slope.  Furthermore, the injected work $W$ is history-dependent as it fluctuates for a given value of the elastic energy $E$.  
This illustrates the multistability of the system, suggesting a complex
energy landscape.

\begin{figure}
\center{\includegraphics[width=.8\linewidth]{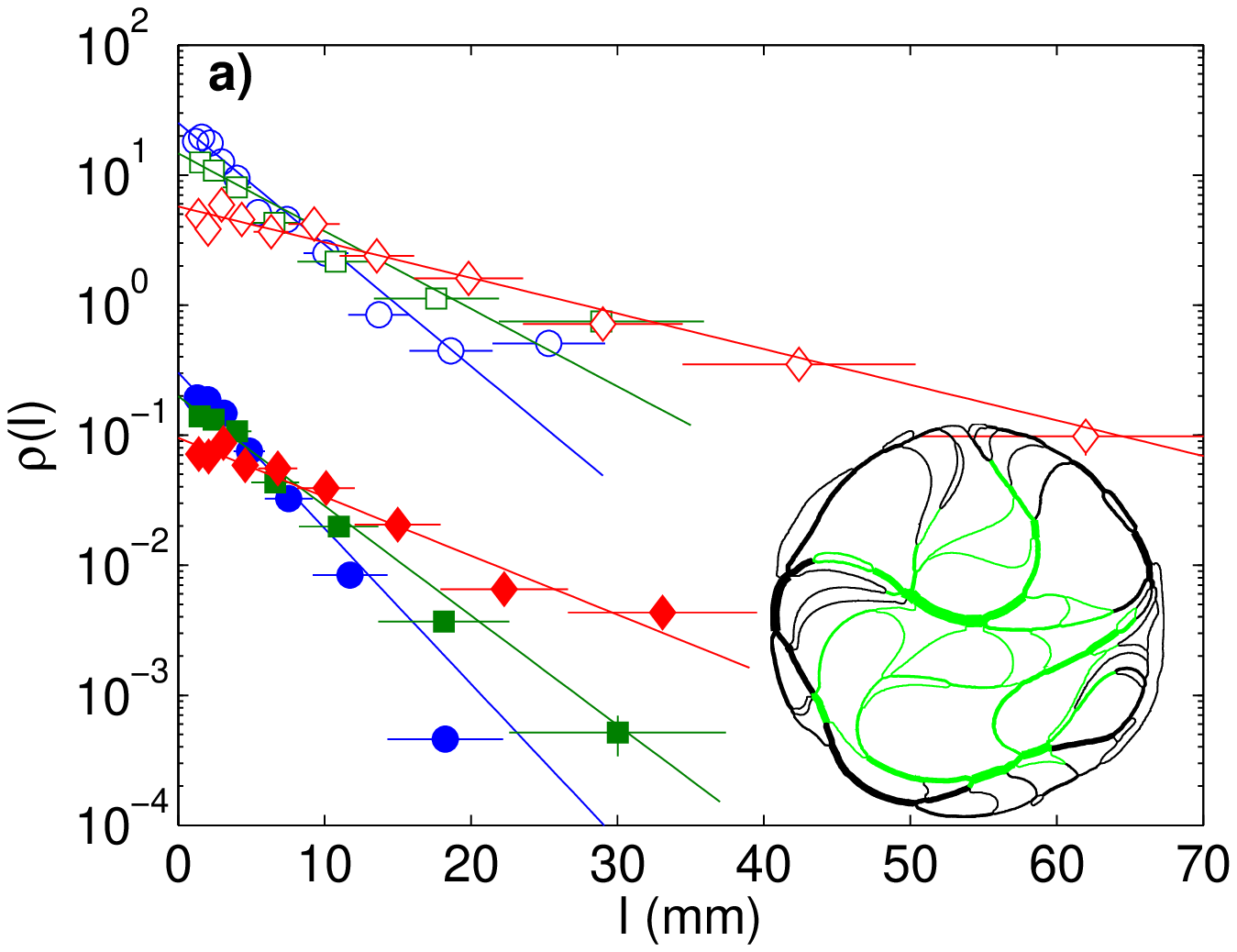}}
\center{\includegraphics[width=.8\linewidth]{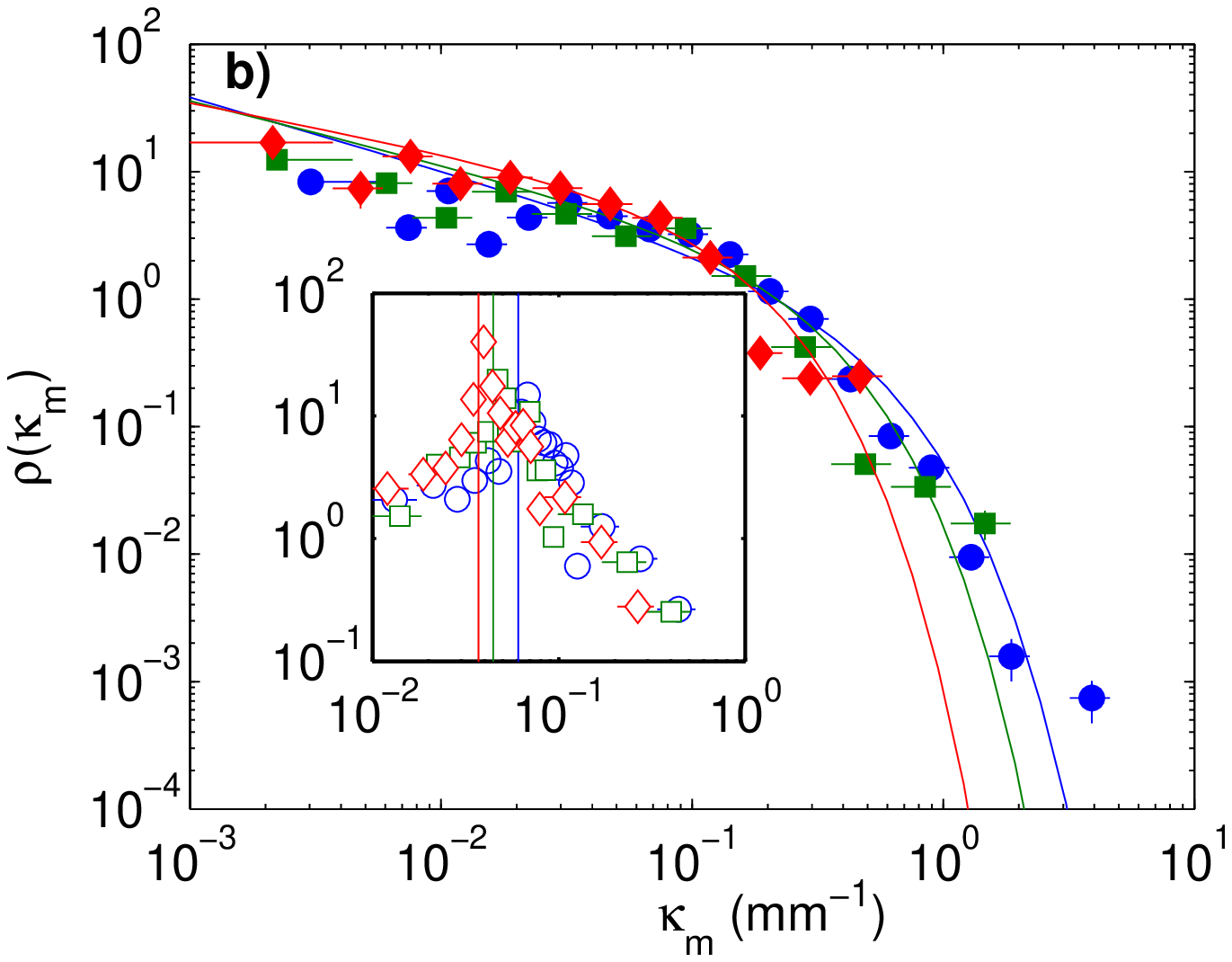}}
\caption{Statistics of the geometrical properties for the three
sets of experiments: $i$ ($\circ$/$\bullet$), $ii$ ($\square$/$\blacksquare$) and $iii$ ($\lozenge$/$\blacklozenge$) respectively in periphery/bulk. \textbf{a)} Experimental pdfs $\rho(\ell)$ of the length $\ell$ of branches and
exponential distributions  $f_\mathrm{E}(\ell)$, Eq.~(\ref{expo}) of the same mean 
as the experimental data. For the periphery $\rho(\ell)$ is multiplied by
$10^2$ for clarity. Means $\langle \ell \rangle=$ $3.3$, $4.5$ mm
for set $i$; $4.6$, $6.8$ mm for set $ii$; and $9.6$, $16$ mm for set
$iii$, respectively for the bulk and the periphery. The inset shows
the two sub-systems: branches in the bulk (green) and at the periphery (dark). \textbf{b)}
Experimental pdfs $\rho(\kappa_m)$ of the mean curvature of branches $\kappa_m$, with same symbols as in
\textbf{a}. For the bulk (main panel), the Gamma distribution 
$f_\mathrm{G}(\kappa_m)$, Eq.~(\ref{gamma}) with the same mean and variance as the experimental data is plotted; its
exponent is $\alpha=0.43$, $0.51$ and $0.62$ and its mean is
$\langle\kappa_m\rangle=0.16$, $0.12$ and $0.08$ mm$^{-1}$, for sets
$i$, $ii$ and $iii$, respectively. For the periphery (inset), the
distributions are peaked at the curvature of the containers $1/R$, shown by
vertical lines.} \label{geometro}
\end{figure}

\section{Statistics of the geometry}

In the following we detail the main statistical properties measured over all realisations of a given set to insure convergence of the statistics. Note that the statistics over one configuration are compatible with ensemble statistics. Within elastic theory of rods, the equilibrium state of a confined rod results from the torque balance of each branch, whereas the interaction between neighbouring
branches is mediated by their extremities where contact forces and
friction come into play~\cite{roman02,boue07}. This fact allows unambiguously to consider branches as the \emph{elementary particles} of the system comparatively to other topologic or geometric decomposition such as contact points or loops which have been used previously~\cite{donato03,stoop08}.

\begin{figure}
\begin{center}
\includegraphics[width=.8\linewidth]{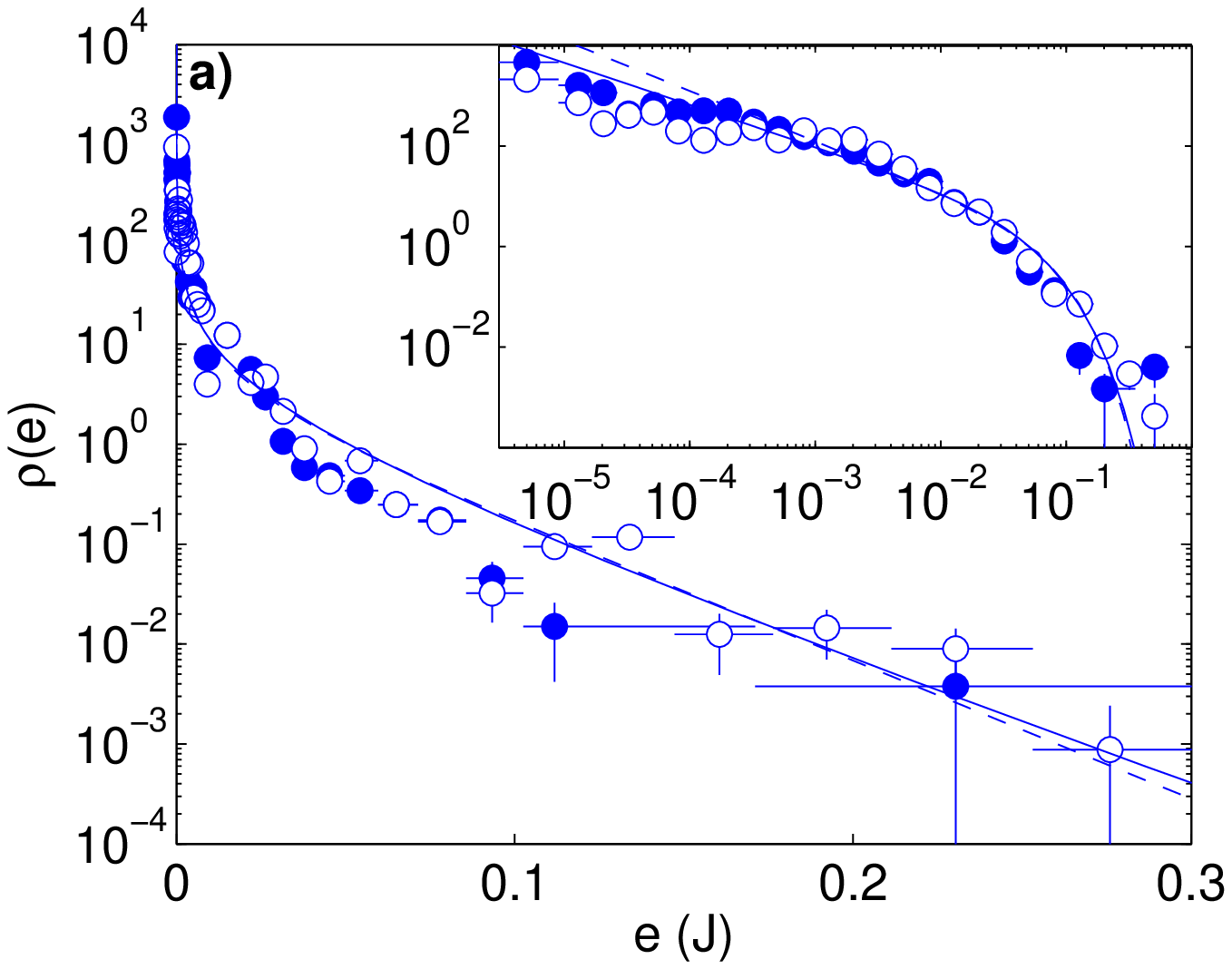}
\includegraphics[width=.8\linewidth]{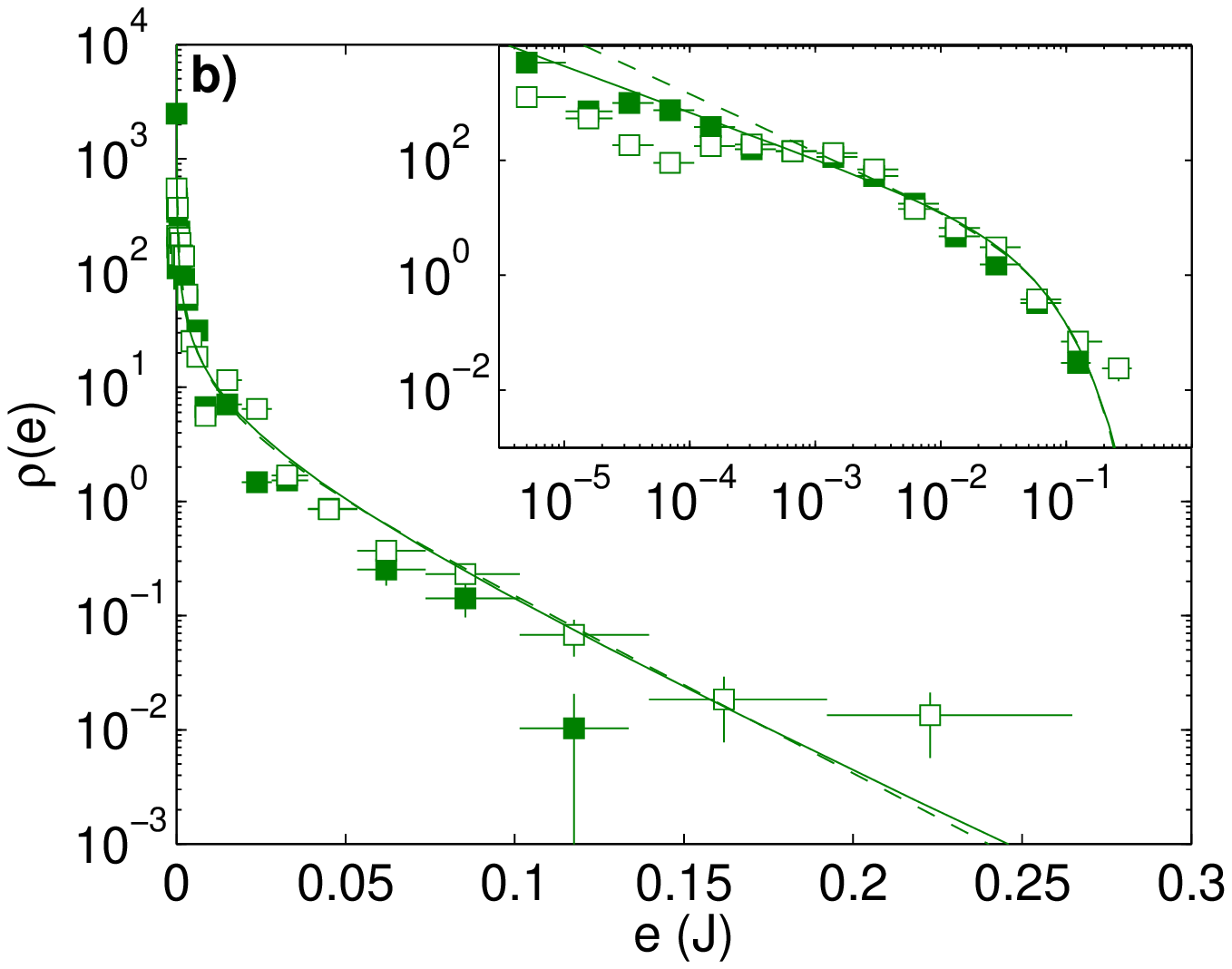}
\includegraphics[width=.8\linewidth]{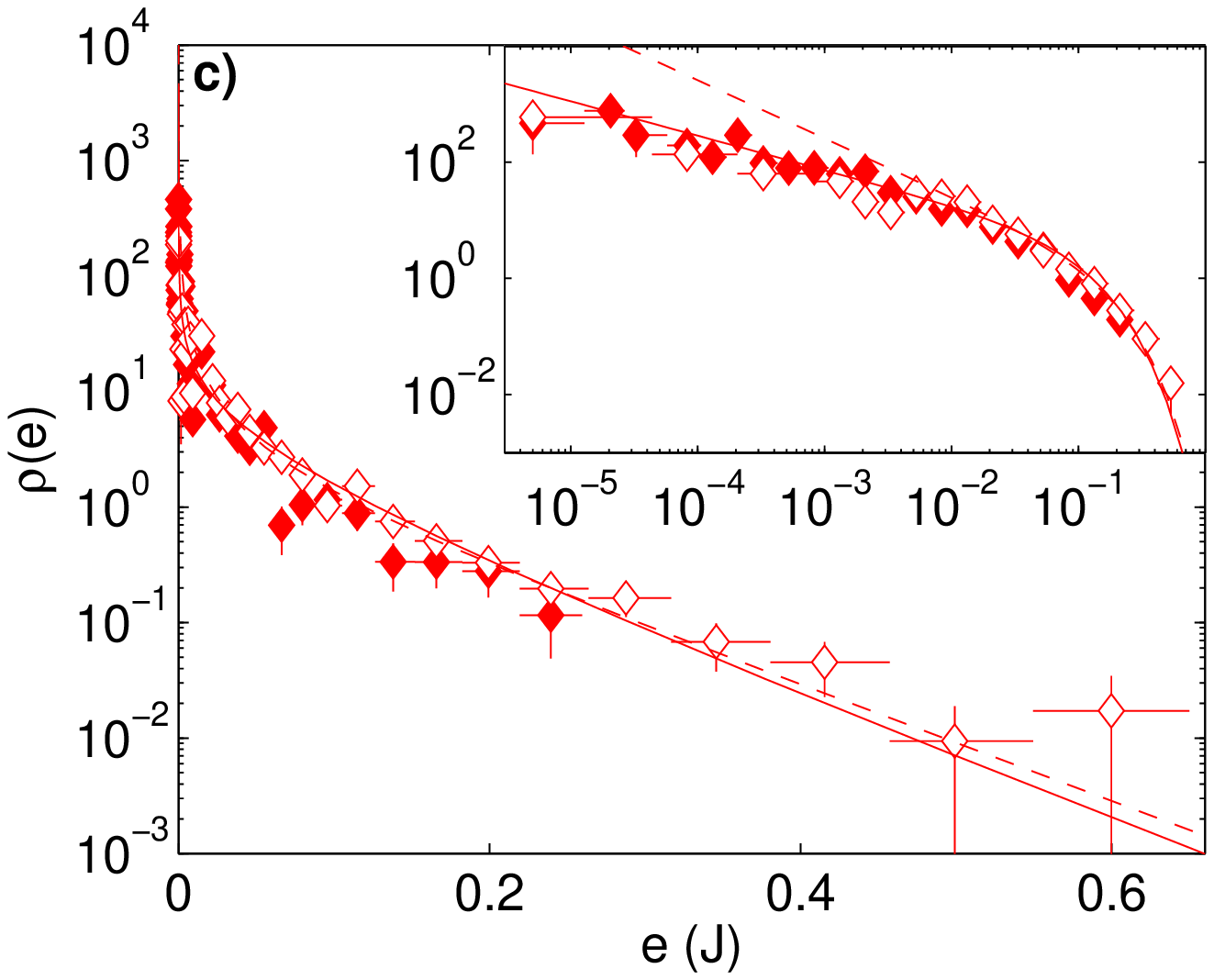}
\end{center}
\caption{Statistics of the energy. Experimental pdfs $\rho(e)$ of the energy $e$  of the
branches  
 for the three sets of
experiments: $i$ (\textbf{a}), $ii$ (\textbf{b}) and $iii$
(\textbf{c}) in log-lin (main panels) and log-log scales (insets). The distributions are given separately for the two
sub-systems: bulk ($\bullet$, $\blacksquare$ and $\blacklozenge$) and periphery ($\circ$, $\square$ and $\lozenge$). The lines are Gamma
distributions Eq.~(\ref{gamma}) with
the same mean and variance as the experimental data. The parameters of the distributions are reported in Table~\ref{tab2}. } \label{energo}
\end{figure}

As the container constrains the curvature of branches in the periphery, it is natural to
 split the system into
two sub-systems (inset in Fig.~\ref{geometro}\textbf{a}): branches
with/without contact with the container, which we
will refer to as \emph{periphery} and \emph{bulk}, respectively. 
The two sub-systems roughly contain the same number of branches ($60\%$ in periphery). 
In the following, we measure probability distribution functions $\rho(x)$ in each sub-system; we compare
these distributions to analytical pdfs $f(x)$ with the same average and variance as experimental data, instead
of direct curve fitting.  The error bars $\delta x$ and $\delta\rho$ of the experimental pdfs $\rho(x)$, shown in Figs.~\ref{geometro} and~\ref{energo}, are given by the bin width $\delta x$ and the estimated standard deviation $\delta\rho=\rho/\sqrt{n}$ of the corresponding histograms $n(x)$. The total number of branches (in periphery and bulk) for a given set of experiments varies between  $\sim2\,10^3$ and $\sim2\,10^4$  (see Table~\ref{tab1}), which allows for accurate statistics. 

Fig.~\ref{geometro}\textbf{a} shows that the lengths $\ell$ of branches 
are well-described by exponential
distributions
\begin{equation}
\label{expo}
f_\mathrm{E}(\ell)=1/\mu\ \exp\left(-{\ell}/{\mu}\right)\textrm{,}
\end{equation}
both in periphery and in bulk.
It appears that the value of the averaged length $\langle \ell
\rangle=\mu$ is significantly larger for branches at the periphery than in the
bulk.

Next, we consider the absolute
value of the average curvature $\kappa_\mathrm{m}$ of each branch
(Fig.~\ref{geometro}\textbf{b}). For the bulk (main panel of Fig.~\ref{geometro}\textbf{b}), the distribution $\rho(\kappa_\mathrm{m})$ is characterised by an exponential tail and a weak power law for small curvatures, which  is
well described by a Gamma law with density
\begin{equation}
\label{gamma}
f_\mathrm{G}^{\alpha,\chi}(\kappa_\mathrm{m})=\frac{(\kappa_\mathrm{m}/\chi)^{\alpha}}{\Gamma(\alpha)\,\kappa_\mathrm{m}}\exp\left(-\frac{\kappa_\mathrm{m}}{\chi}\right)
\textrm{,}
\end{equation}
where $\Gamma$ stands for Euler's Gamma function. In contrast, for the periphery (inset of Fig.~\ref{geometro}\textbf{b}), the distribution  $\rho(\kappa_\mathrm{m})$ is peaked around the value $1/R$ given by the container. Thus, the geometrical properties of the periphery and the bulk are significantly different.

\section{Statistics of the energy}

Each configuration is at mechanical equilibrium, so that any branch can be characterised by its elastic
energy. 
The energy $e$ of the branch corresponds to that of an angular sector on the sheet and is calculated using Eq.~(\ref{eqformenerg}), with limits of integration $t\in(0,\ell)$, where $\ell$ is the branch length. 
Surprisingly, the probability distribution functions $\rho(e)$ in periphery and in bulk coincide, as shown in Fig.~\ref{energo} \textbf{a}, \textbf{b} and \textbf{c} for the three sets of experiments $i$, $ii$ and $iii$ respectively. Despite the heterogeneous geometry of the branches, which contributes to their energy through length and curvature,  the energy is homogeneous inside the whole system. These distributions 
are characterised by a power-law divergence at small values and by exponential tails, as shown by the log-log and log-lin scales in inset and main panels of Fig.~\ref{energo}. Thus, it is natural to compare them to Gamma distributions. Indeed, they are well-described by Gamma laws $f_\mathrm{G}^{\alpha_e,\chi_e}(e)$ as given by Eq.~(\ref{gamma}), with the same average and variance as the experimental data (see Fig.~\ref{energo}).  However, we found exponents $\alpha_e<1$ which are not trivially interpretable as a number of effective subsystems,
in contrast with the cases where $\alpha_e>1$~\cite{sultan06,lechenault06}. This point will be discussed below.

Here we emphasise that the shape of the distributions of energy is insensitive neither to the overall multiplicative factor $\log(r/R_c)$ of Eq.~(\ref{eqformenerg}) nor to the plastic threshold $\kappa_c$. The former amounts to a normalisation of the average energy of a given set, while the value of the latter does not change the statistics since it affects only a few branches.

\section{Discussion}

We investigated the close packing of elastic sheets in a quasi two-dimensional experimental 
setup allowing the statistical study of the geometry and the energy of the resulting configurations. 
These quantities are broadly distributed, suggesting a complex energy landscape. We identified
branches as natural \emph{elementary particles}: the shape of a branch is completely prescribed
by its length and boundary conditions. The interaction between branches is mediated by the contact
forces at their extremities, which is reminiscent of granular packing. The presence of the rigid container led  
us to split the system into two sub-systems: periphery and bulk. It turns out that the energy of branches is
the only quantity which is identically distributed in the two sub-systems, even though the geometrical properties differ. 
This homogeneity of the distributions of energy is our central result. This property might be an indication of thermal equilibration. Future work should address this important question.

Moreover, the energy distributions of the different sets of experiments are characterised by an exponential tail that is reminiscent of Boltzmann distributions. 
Consequently, the distributions of energy allow to define effective temperatures for each set of experiments: the mean energy per branch $\langle e\rangle$ and the characteristic energy given by the exponential tails $\chi_e$. The effective temperatures are ordered as $\langle e\rangle < \chi_e$ for each set of experiments (Table~\ref{tab2}); the sets of temperatures are close for the two sets of experiments with the same thickness $h$ and bending stiffness $B$ ($i$ and $ii$), whereas these correspond to very different packing ratios ($P=0.11$ and $0.06$). This suggests that the bending stiffness $B$ might be relevant
for the value of the effective temperatures. However, more work is needed with this respect because of the inaccurate estimation of the overall logarithmic multiplicative factor in Eq.~(\ref{eqformenerg}).

As stated above and shown in Table~\ref{tab2}, the exponents $\alpha_e<1$ are not trivialy interpretable as a number of effective subsystems, which makes difficult the physical interpretation of the Gamma distributions. Nevertheless, another distribution function exhibiting a power-law at small values and an exponential tail is provided by Bose-Einstein statistics
\begin{equation}
\label{bose} f_\mathrm{BE}(e) = \frac{g(e)}{\exp\left(\beta e \right)-1}
 \textrm{,}
\end{equation}
when the chemical potential vanishes. Here $g(e)$ is the density of states. In the case of noninteracting bosons $g(e)\sim e^{(d-2)/2}$ where $d$ is the space dimension, which can lead to a divergent behaviour of the distribution at small energies. Thus the distributions measured here could be interpreted as obeying a Bose-Einstein statistics with a power-law $g(e)$. A rationale would be as follows: many branches may be in the same state (when belonging to the same stack) as bosons; the number of branches is unprescribed so that the `chemical potential' is zero.

Thermal homogeneity suggests a description of the system in terms of statistical physics. However, our system is obviously not ergodic, as it must be driven by injecting work in order to explore the phase space. This driving has some similarities with the slow shearing of colloidal glasses~\cite{berthier02} or granular materials~\cite{makse02,song05}; however, it is not stationary and restricts the accessible phase space at each reconfiguration of the sheet. As in other glassy systems, two different time scales characterise the dynamics: a very slow one associated with the driving and a quick one corresponding to the reconfiguration to local mechanical equilibrium. 
Finally, further experimental and theoretical work is needed to explain our observations  and to confirm our interpretations. Can one predict the distributions from first principles? 
 How universal are these distributions? What controls the effective temperatures measured here?  

\begin{table}
\begin{center}
  \begin{tabular}{l||l|l|l}
  & $\langle e\rangle$ (mJ)& $\chi_{e}$ (mJ)&$\alpha_e$   \\ \hline
$1$   & 6.5  & 39 &0.16     \\
$2$  & 6.6  & 35  &0.23   \\
$3$ & 37 & 90 &0.41
  \end{tabular}
\end{center}
  \caption{Effective temperatures: $\langle e \rangle$  is the mean energy per \emph{elementary particle}, i.e. per branch; $\chi_e$ and $\alpha_e$ are given by the tail and the exponent of the Gamma distribution of energy in Fig.~\ref{energo}. }
\label{tab2}
\end{table}

\acknowledgments
We are grateful to G. Angot and J. Da~Silva-Quintas for their experimental and technical help.
This study was supported by the EU through the NEST MechPlant project. LPS is associated with the universities of Paris VI and VII.

\end{document}